\documentclass{pasj01}

\input{pasjsup.sty}

\newcounter{author}
\def\authorcount#1#2{\refstepcounter{author}\label{#1}
\altaffiltext{\ref{#1}}{#2}}

\begin{document}
\SetRunningHead{Y. Wakamatsu et al.}{Title}

\Received{201X/XX/XX}
\Accepted{201X/XX/XX}

\title{ASASSN-16eg: New candidate of long-period WZ Sge-type dwarf nova}

\author{Yasuyuki~\textsc{Wakamatsu},\altaffilmark{\ref{affil:AA}*}
Keisuke~\textsc{Isogai},\altaffilmark{\ref{affil:AA}}
Mariko~\textsc{Kimura},\altaffilmark{\ref{affil:AA}}
Taichi~\textsc{Kato},\altaffilmark{\ref{affil:AA}}
Tonny~\textsc{Vanmunster},\altaffilmark{\ref{affil:AB}}
Geoff~\textsc{Stone},\altaffilmark{\ref{affil:AC}}
Tam\'as~\textsc{Tordai},\altaffilmark{\ref{affil:AD}}
Michael~\textsc{Richmond},\altaffilmark{\ref{affil:AE}}
Ian~\textsc{Miller},\altaffilmark{\ref{affil:AF}}
Arto~\textsc{Oksanen},\altaffilmark{\ref{affil:AG}}
Hiroshi~\textsc{Itoh},\altaffilmark{\ref{affil:AH}}
Hidehiko~\textsc{Akazawa},\altaffilmark{\ref{affil:AI}}
Seiichiro~\textsc{Kiyota},\altaffilmark{\ref{affil:AJ}}
Enrique de~\textsc{Miguel},\altaffilmark{\ref{affil:AK},\ref{affil:AL}}
Elena P.~\textsc{Pavlenko},\altaffilmark{\ref{affil:AM}}
Kirill A.~\textsc{Antonyuk},\altaffilmark{\ref{affil:AM}}
Oksana I.~\textsc{Antonyuk},\altaffilmark{\ref{affil:AM}}
Vitaly V.~\textsc{Neustroev},\altaffilmark{\ref{affil:AN},\ref{affil:AO}}
George~\textsc{Sjoberg},\altaffilmark{\ref{affil:AP},\ref{affil:AQ}}
Pavol A.~\textsc{Dubovsky},\altaffilmark{\ref{affil:AR}}
Roger D.~\textsc{Pickard},\altaffilmark{\ref{affil:AS},\ref{affil:AT}} and
Daisaku~\textsc{Nogami}\altaffilmark{\ref{affil:AA}}
}

\authorcount{affil:AA}{
Department of Astronomy, Kyoto University, Kyoto 606-8502, Japan}
\authorcount{affil:AB}{
Center for Backyard Astrophysics Belgium, Walhostraat 1A, B-3401 Landen, Belgium}
\authorcount{affil:AC}{
Sierra Remote Observatories, 44325 Alder Heights Road, Auberry, CA 93602, USA}
\authorcount{affil:AD}{
Polaris Observatory, Hungarian Astronomical Association, Laborc utca 2/c, 1037 Budapest, Hungary}
\authorcount{affil:AE}{
Physics Department, Rochester Institute of Technology, Rochester, New York 14623, USA}
\authorcount{affil:AF}{
Furzehill House, Ilston, Swansea, SA2 7LE, UK}
\authorcount{affil:AG}{
Hankasalmi observatory, Jyvaskylan Sirius ry, Murtoistentie 116, FI-41500 Hankasalmi, Finland}
\authorcount{affil:AH}{
Variable Star Observers League in Japan (VSOLJ), 1001-105 Nishiterakata, Hachioji, Tokyo 192-0153, Japan}
\authorcount{affil:AI}{
Department of Biosphere-Geosphere System Science, Faculty of Informatics, Okayama University of Science, 1-1 Ridai-cho, Okayama, Okayama 700-0005, Japan}
\authorcount{affil:AJ}{
VSOLJ, 7-1 Kitahatsutomi, Kamagaya, Chiba 273-0126, Japan}
\authorcount{affil:AK}{
Departamento de Ciencias Integradas, Facultad de Ciencias Experimentales, Universidad de Huelva, 21071 Huelva, Spain}
\authorcount{affil:AL}{
Center for Backyard Astrophysics, Observatorio del CIECEM, Parque Dunar, Matalasca\~nas, 21760 Almonte, Huelva, Spain}
\authorcount{affil:AM}{
Federal State Budget Scientific Institution “Crimean Astrophysical Observatory of RAS”, Nauchny, 298409, Republic of Crimea}
\authorcount{affil:AN}{
Finnish Centre for Astronomy with ESO (FINCA), University of Turku, V\"{a}is\"{a}l\"{a}ntie 20, FIN-21500 Piikki\"{o}, Finland}
\authorcount{affil:AO}{
Astronomy research unit, PO Box 3000, FIN-90014 University of Oulu, Finland}
\authorcount{affil:AP}{
The George-Elma Observatory, New Mexico Skies, 9 Contentment
Crest, \#182, Mayhill, NM 88339, USA}
\authorcount{affil:AQ}{
American Association of Variable Star Observers, 49 Bay State Rd., Cambridge, MA
02138, USA}
\authorcount{affil:AR}{
Vihorlat Observatory, Mierova 4, 06601 Humenne, Slovakia}
\authorcount{affil:AS}{
The British Astronomical Association, Variable Star Section (BAA VSS), Burlington House, Piccadilly, London, W1J 0DU, UK}
\authorcount{affil:AT}{
3 The Birches, Shobdon, Leominster, Herefordshire, HR6 9NG, UK}

\email{$^*$wakamatsu@kusastro.kyoto-u.ac.jp}


\maketitle

\begin{abstract}
We report on our photometric observations of the 2016 superoutburst of ASASSN-16eg.
This object showed a WZ Sge-type superoutburst with prominent early superhumps with a period of 0.075478(8) d and a post-superoutburst rebrightening.
During the superoutburst plateau, it showed ordinary superhumps with a period of 0.077880(3) d and a period derivative of 10.6(1.1) $\times$ 10$^{-5}$ in stage B.
The orbital period ($P_{\rm orb}$), which is almost identical with the period of early superhumps, is exceptionally long for a WZ Sge-type dwarf nova.
The mass ratio ($q$ = $M_2/M_1$) estimated from the period of developing (stage A) superhumps is 0.166(2), which is also very large for a WZ Sge-type dwarf nova.
This suggests that the 2:1 resonance can be reached in such high-$q$ systems, contrary to our expectation.
Such conditions are considered to be achieved if the mass-transfer rate is much lower than those in typical SU UMa-type dwarf novae that have comparable orbital periods to ASASSN-16eg and a resultant accumulation of a large amount of matter on the disk is realized at the onset of an outburst.
We examined other candidates of long-period WZ Sge-type dwarf novae for their supercycles, which are considered to reflect the mass-transfer rate, and found that V1251 Cyg and RZ Leo have longer supercycles than those of other WZ Sge-type dwarf novae.
This result indicates that these long-period objects including ASASSN-16eg have a low mass-transfer rate in comparison to other WZ Sge-type dwarf novae.
\end{abstract}

\section{Introduction}
\label{introduction}
Cataclysmic variables (CVs) are close binary systems composed of a white dwarf (WD) primary and a secondary that transfers mass to the primary.
The secondary fills its Roche lobe and its overflowing matter pours onto the primary through the inner Lagrangian point $L_1$.
Dwarf novae (DNe) are a subclass of CVs and have a property of recurrent outbursts with typically 2--5 mag brightening.
The outburst lasts for days or weeks.
It is considered that the outburst results from a sudden release of gravitational energy which is caused by a rapid increase of the mass accretion rate on the primary by the thermal instability in the disk \citep{osa74DNmodel}.

SU UMa-type DNe are a subclass of DNe characterized by occasional superoutbursts, which have longer durations than normal outbursts with superhumps.
Superhumps are variations of small amplitudes typically of 0.1--0.5 mag and are considered to be a result of the tidal instability that is triggered when the outer disk reaches the 3:1 resonance radius during the outburst \citep{whi88tidal, osa89suuma, lub91SHa, lub91SHb, hir90SHexcess}.
\citet{Pdot} proposed that the superoutburst is divided into three distinct stages by a variation of the superhump period ($P_{\rm SH}$): stage A has a longer superhump period, in stage B the superhump period systematically varies, and stage C is a final stage of superoutburst and has a shorter superhump period.
The amplitude of the superhumps grows during stage A and then decreases during stage B.
Intervals between the superoutbursts (supercycles) are typically several hundred days.

WZ Sge-type DNe are a subclass of the SU UMa-type and show mainly superoutbursts and rarely normal outbursts.
WZ Sge-type DNe are characterized by the large amplitude, long duration superoutbursts and in some cases the existence of post-superoutburst brightenings, which are called rebrightenings \citep[for more detail]{kat15wzsge}.
WZ Sge-type DNe are also characterized by double-wave small variations of magnitude, which are called early superhumps and have almost the same period as the orbital period, before a growth of the stage A superhumps \citep{kat02wzsgeESH, Pdot6, ish01rzleo, ish02wzsgeletter}.
Although the historical classifications of WZ Sge-type DNe were mainly based on the amplitude of superoutbursts (e.g., tremendous outburst amplitude dwarf novae or TOADs \citep{how95TOAD}), the presence of early superhumps is now considered to be a key criterion for classification of WZ Sge-type DNe \citep{kat15wzsge}\footnote{The definitions of WZ Sge-type DNe by American Association of Variable Star Observers (AAVSO) International Variable Star Index (VSX) are (i)unusual long supercycle, (ii)existence of early superhumps, (iii)existence of rebrightenings, (iv)orbital periods with range of 0.05--0.08 d and (v)outburst amplitudes larger than $\sim$7 mag.}.
It is considered that early superhumps are caused when the outer edge of the disk reaches the 2:1 resonance radius during the outburst.
The 2:1 resonance is considered to suppress the deformation of the disk caused by the 3:1 resonance, and the eccentricity change due to the 3:1 resonance grows when the outer edge of the disk falls below the 2:1 resonance radius \citep{osa02wzsgehump,lub91SHa}.
As a consequence, early superhumps are observed in an early stage of the superoutburst and then ordinary superhumps appear subsequently instead of the early superhumps.
In order to reach the 2:1 resonance radius, it is considered that a mass ratio $q$ = $M_2/M_1$ ($M_1$ and $M_2$ represent the mass of the primary and secondary, respectively) should be extremely low.
\citet{osa02wzsgehump} proposed that the outer edge of the disk can reach the 2:1 resonance radius in the low mass-ratio systems with $q<$ 0.08.
Indeed, WZ Sge-type DNe have extremely small mass ratios, which are typically 0.06--0.08, and also have short orbital periods, $P_{\rm orb}$, which are around 0.054--0.056 d \citep{kat15wzsge}.
The mass-transfer rates from the secondary in WZ Sge-type DNe are very small in comparison with typical SU UMa-type DNe and the supercycles are extremely long.
The supercycles are typically a few years or decades \citep{kat15wzsge}.
There are, however, some unusual objects which show superoutbursts with WZ Sge-type features, i.e., large amplitudes of superoutbursts, the existence of rebrightenings or in some systems, the existence of double-wave modulations similar to early superhumps, although  they have longer orbital periods than those of other WZ Sge-type DNe.
These long-period objects are classified as WZ Sge-type DNe in \citet{kat15wzsge} based on observational features.

According to the standard evolutionary theory of CVs, a binary separation becomes shorter because of a loss of the total angular momentum due to the magnetic braking and/or gravitational radiation.
If the mass transfer continues, the secondary becomes degenerate at a certain orbital period and then the binary evolves as its separation becomes wider.
There is, therefore, a minimum orbital period of CVs, and the binaries passing the period minimum are called period bouncers \citep[and references therein]{kol99CVperiodminimum,kni11CVdonor}.
As CVs evolve, the mass-transfer rate becomes lower and the supercycles become longer.
Therefore, the mass-transfer rates in WZ Sge-type DNe are considered to be smaller than those in SU UMa-type DNe \citep{osa02wzsgehump}.

In this paper, we present observations of the ASASSN-16eg.
The superoutburst of ASASSN-16eg was detected on 2016 April 9 by All-Sky Automated Survey for Supernovae (ASAS-SN; \citet{ASASSN}) and the magnitude was $V$ = 14.4 at the time of detection.
The coordinates of this object are RA: 17$^{\rm h}$26$^{\rm m}$10$^{\rm s}$.3213 and Dec: +42$^\circ$20$^\prime$02$^{\prime \prime}$.660 (J2000.0) in {\it Gaia} Data Release 1 \citep[for more detail about {\it Gaia} DR1]{GaiaDR1}.
There is a quiescent counterpart of G=19.394 in {\it Gaia} DR1.
ASASSN-16eg was classified in WZ Sge-type DN since this object showed apparently clear double-wave modulations having properties of early superhumps and rebrightening, although its orbital period is particularly long.
We found that ASASSN-16eg has a considerably large mass ratio which is far beyond the upper limit of the mass ratio that the outer edge of the disk is supposed to be able to reach the 2:1 resonance radius \citep{osa02wzsgehump}.
We considered why this object showed a superoutburst that stems from 2:1 resonance despite of its large mass ratio by comparing with other long-period objects.
In section \ref{observation}, we describe the details of our observations and the methods of analyses. In section \ref{result}, we present the results of our observations. In section \ref{discussion}, we discuss the results.

\begin{figure*}
\begin{center}
\FigureFile(170mm,100mm){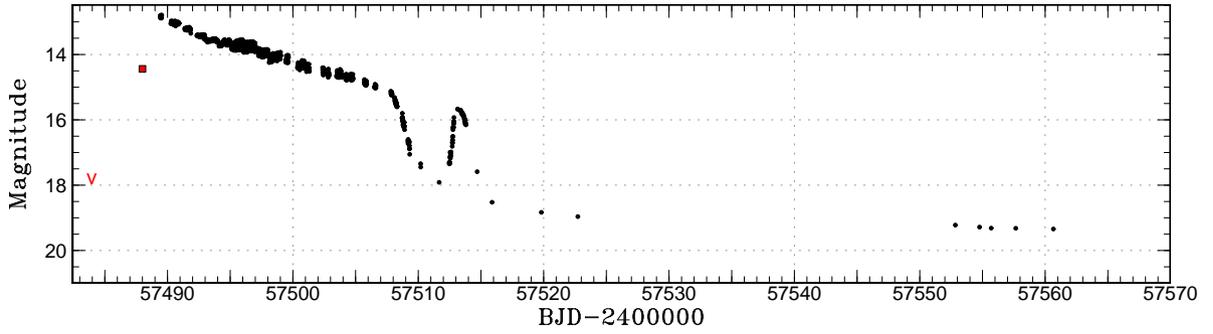}
\end{center}
\caption{The overall light curve of the 2016 superoutburst of ASASSN-16eg. The filled square and V-shaped sign represent an observational point and upper limit by ASAS-SN, respectively.}
\label{lightcurve}
\end{figure*}

\section{Observation and Analysis}
\label{observation}
Our time-resolved CCD photometry of the superoutburst of ASASSN-16eg was carried out by the Variable Star Network (VSNET) collaborations \citep{VSNET}.
Logs of our photometric observations are in table S1.
All of the observation times were described in barycentric Julian date (BJD).
We added a constant to each observer's magnitude data to adjust the difference in the zero-point.
We used the phase dispersion minimization (PDM) method \citep{PDM} for period analyses.
The 1$\sigma$ error of the best estimated period by the PDM method was determined by the methods in \citet{fer89error} and \citet{Pdot2}.
We subtracted the global trend of the light curve by subtracting a smoothed light curve obtained by locally weighted polynomial regression (LOWESS: \cite{LOWESS}) before making the period analyses.
We used $O-C$ diagrams, from which we can derive the slight variation of the superhump period (see, e.g., \cite{ste05OCdiagram}).

\section{Result}
\label{result}

\begin{figure}
\begin{center}
\FigureFile(80mm,100mm){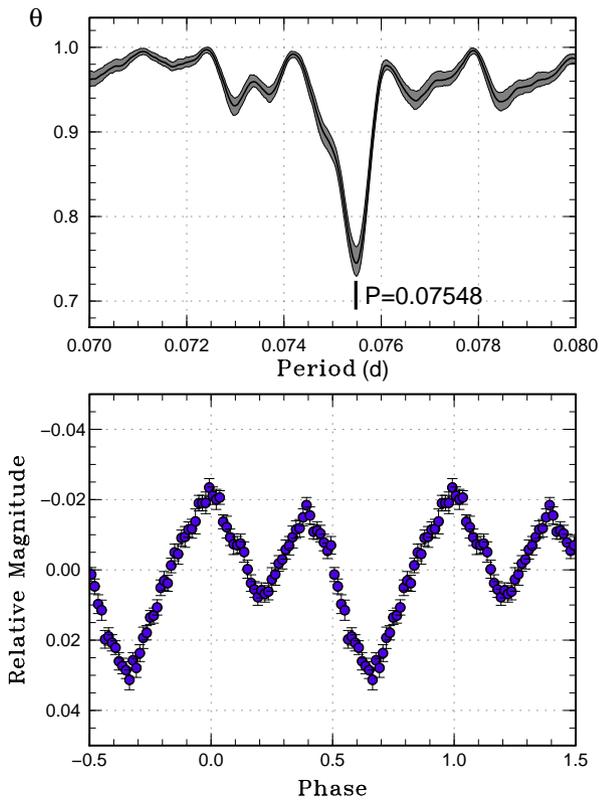}
\end{center}
\caption{Upper panel: $\theta$-diagram of our PDM analysis of early superhumps of ASASSN-16eg (BJD 2457489.4--2457494.0). The gray area represents the 1$\sigma$ error of the best estimated period by the PDM method. Lower panel: Phase-averaged profile of early superhumps.}
\label{a16egpdmearly}
\end{figure}

\subsection{Overall light curve}
Figure \ref{lightcurve} shows the overall light curve of the superoutburst of ASASSN-16eg.
The observation was started on BJD 2457488.
The superoutburst lasted about 20 d during BJD 2457489--2457508 with a slow decline of the brightness, and then rapidly faded.
There was a single rebrightening during BJD 2457512--2457516.
After the rebrightening, the magnitude declined to around V=19.5 and ASASSN-16eg seemed to be in a quiescent state.

\subsection{Early superhumps}
We regarded the variations recorded in BJD 2457489.4--2457494.0 as early superhumps based on the double-wave variation and the variation of the superhump period.
Figure \ref{a16egpdmearly} shows the result of PDM analysis of early superhumps (upper panel) and the mean profile (lower panel) of ASASSN-16eg.
Double-wave variations characterized as early superhumps are clearly seen.
We found the period of early superhumps to be 0.075478(8) d.

\begin{figure}
\begin{center}
\FigureFile(85mm,120mm){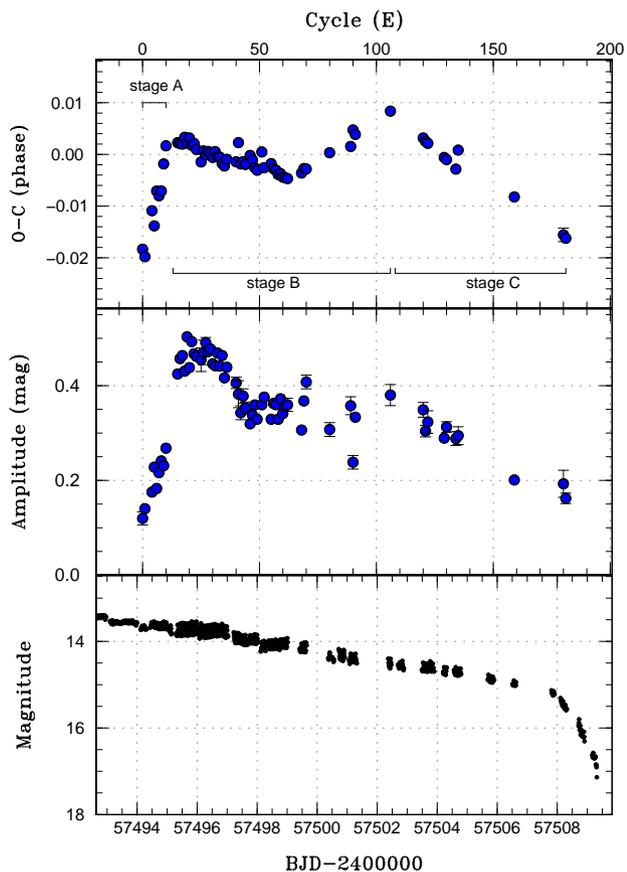}
\end{center}
\caption{Upper panel: The $O-C$ curve of ASASSN-16eg during BJD 2457494--2457508. We used an ephemeris of BJD 2457496.4415+0.0779132$E$ for drawing this figure. Middle panel: The amplitude of superhumps. Lower panel: The light curve. The horizontal axis in units of BJD and cycle number is common to all of these panels.}
\label{a16egOCcurve}
\end{figure}

\begin{figure}
\begin{center}
\FigureFile(80mm,100mm){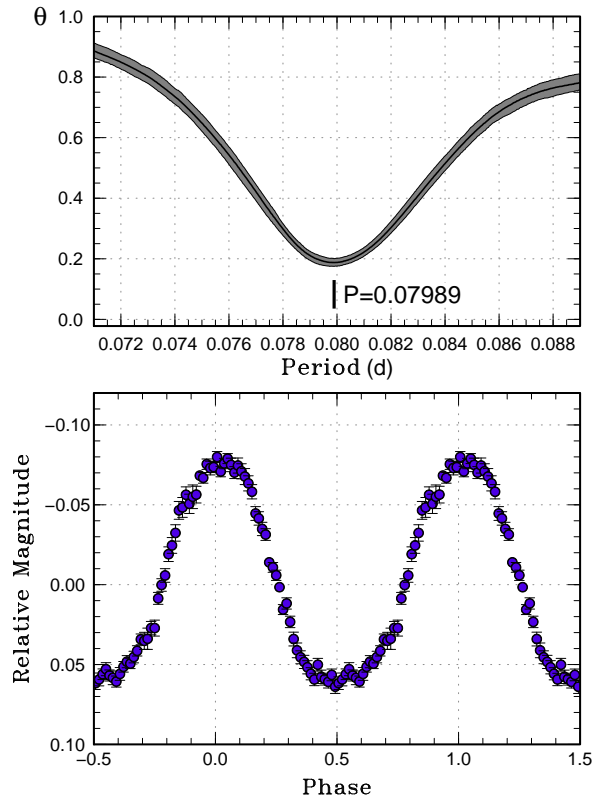}
\end{center}
\caption{Upper panel: $\theta$-diagram of our PDM analysis of stage A superhumps of ASASSN-16eg (BJD 2457494.1-2457495.1). The gray area represents the 1$\sigma$ error of the best estimated period by the PDM method. Lower panel: Phase-averaged profile of stage A superhumps.}
\label{a16egpdmA}
\end{figure}

\subsection{Ordinary superhumps}
Figure \ref{a16egOCcurve} shows the $O-C$ curve (upper panel), the amplitude of the superhumps (middle panel) and the light curve (lower panel) of ASASSN-16eg during BJD 2457494--2457508.
We determined the times of maxima of ordinary superhumps in the same way as in \citet{Pdot}.
The resultant times are listed in table S2.
We regarded BJD 2457494.1--2457495.1 (0 $\leq E \leq$ 10) as stage A, BJD 2457495.2--2457502.9 (15 $\leq E \leq$ 106) as stage B, and BJD 2457503.4--2457508.4 (120 $\leq E \leq$ 181) as stage C superhumps from the variation of the superhump period and the amplitude of superhumps.

Figure \ref{a16egpdmA} shows the result of PDM analysis of stage A superhumps (upper panel) and the mean profile (lower panel).
We found the stage A superhump period to be $P_{\rm stA}$ = 0.07989(4) d.
We also found the stage B and stage C superhump period to be 0.077880(3) d and 0.077589(7) d, respectively.
$P_{\rm dot} (\equiv \dot P_{\rm sh}/P_{\rm sh})$, which is a derivative of the superhump period during stage B, was 10.4(0.8) $\times$ 10$^{-5}$.

\section{Discussion}
\label{discussion}

\subsection{Particularly long orbital period and large mass ratio}
\label{massratio_and_period}
As the period of early superhumps is considered to be almost equal to the orbital period \citep{Pdot6}, we estimated the orbital period of ASASSN-16eg to be $P_{\rm orb}$ = 0.075478(8) d.
This value is particularly large compared with that of other WZ Sge-type DNe, which are concentrated around 0.054--0.056 d \citep{kat15wzsge}.
Such long orbital period indicates two possibilities either ASASSN-16eg is in an earlier stage of CV evolution than other WZ Sge-type DNe or ASASSN-16eg is in a final stage of CV evolution as a period bouncer.
We excluded, however, the latter possibility because of its large mass ratio as discussed below.
There are some objects suspected as WZ Sge-type DNe having long orbital periods like ASASSN-16eg.
ASASSN-16eg may be a new candidate of these long-period objects (see subsection \ref{longporb}).

We estimated the mass ratio of ASASSN-16eg from the fractional superhump-period excess for the 3:1 resonance, $\varepsilon^*$ = 1 $-P_{\rm orb}/P_{\rm stA}$, in the same way as proposed in \citet{kat13qfromstageA}.
We estimated $\varepsilon^*$ = 0.0552(6) and then found the mass ratio to be $q$ = 0.166(2).
This value is considerably large for other WZ Sge-type DNe, which are around 0.06--0.08 \citep{kat15wzsge}, and thus the mass ratio of ASASSN-16eg is twice or three times as large as these typical values.

Both the orbital period and the mass ratio of ASASSN-16eg are similar to those of SU UMa-type DNe rather than WZ Sge-type ones (see figure \ref{evol-track}).
However, we note that the period of early superhumps is not exactly but almost equal to the orbital period.
\citet{ish02wzsgeletter} showed that the period of early superhumps of WZ Sge is 0.05\% shorter than the orbital period.
\citet{Pdot6} showed that the differences between the periods of early superhumps and the orbital periods in well-studied WZ Sge-type DNe are very small and periods of early superhumps can be used as approximate orbital periods with an accuracy of 0.1\%, and they also proposed that we can derive the orbital period from the period of early superhumps by assuming fractional excess of early superhumps, $\varepsilon$, of $-$0.05\% if more accuracy is needed.
Considering this difference in ASASSN-16eg, we obtained an improved orbital period of 0.07544026(8) d and then estimated the mass ratio to be 0.167(2) by using the method proposed in \citet{kat13qfromstageA}.
This value is very close to that we estimated from the period of early superhumps, and thus the difference between the period of early superhumps and the orbital period is considered to be not important for the estimation of the mass ratio.

\begin{table*}
\caption{Candidates of long-period objects showing WZ Sge-type superoutbursts}
  \begin{tabular}{p{2.6cm}p{1.6cm}p{1.6cm}p{1.2cm}p{3.4cm}p{2cm}p{1.6cm}}
  \hline
  Object & $P_{\rm orb}$ (d) & $P_{\rm stA}$ (d) & $q$ & Superoutbursts & Supercycle (yr) & References\\
  \hline
  V1251 Cyg & -- & 0.07616(3) & -- & 1963, 1991, 1994, 1997, 2008\commenta & 3--28 & 1, 2, 3, 4\\
  RZ Leo & 0.07626(7) & 0.08072(5) & 0.165(6) & 1918, 1935, 1952, 1976, 1984, 2000\commenta, 2006, 2016 & 6--24 & 5, 6, 7, 8, 9\\
  BC UMa & 0.06251(5) & 0.06476(7) & 0.096(6) & 1960, 1962, 1982, 1990, 1992, 1994, 2000, 2003\commenta, 2009 & 2--20 & 1, 6, 10, 11, 12, 13, 14\\
  MASTER J004527 & -- & 0.08136(7) & -- & 2013\commenta & -- & 15\\
  QY Per & -- & -- & -- & 1999\commenta, 2005, 2015 & 6--10 & 1, 5\\
  ASASSN-16eg & 0.075478(8) & 0.07989(4) & 0.166(2) & 2016\commenta & -- & this paper\\
  \hline
  \end{tabular}
  \begin{tabular}{p{16.6cm}}
	References: 1. \citet{Pdot}; 2. \citet{web66v1251cyg}; 3. \citet{wen91v1251cyg}; 4. \citet{kat95v1251cyg}; 5. \citet{Pdot8}; 6. \citet{kat15wzsge}; 7. \citet{wol19rzleo}; 8. \citet{ric85rzleo}; 9. \citet{ish01rzleo}; 10. \citet{Pdot2}; 11. \citet{rom64bcuma}; 12. \citet{kun98bcuma}; 13. \citet{boy03bcuma}; 14. \citet{mae07bcuma}; 15. \citet{Pdot6}
  \end{tabular}
  \begin{tabular}{p{16.6cm}}
  \commenta Superoutbursts that we reanalyzed in this paper.
  \end{tabular}
  \label{tab:longperiod}
\end{table*}

\subsection{Conditions of the 2:1 resonance}
\label{2:1resonance}
\citet{osa02wzsgehump} proposed that the outer edge of the disk cannot reach the 2:1 resonance radius in the systems with $q>$ 0.08.
This upper limit of $q$ for the 2:1 resonance may not be a rigid value since this extension of the disk radius was derived under an assumption of angular momentum conservation of the steady hot disk.
If accretion onto the primary proceeds during the outburst, the outer edge of the disk may expand beyond this radius until the resonance radius or the tidal truncation radius stops its expansion.
Accretion of a large amount of matter onto the primary will lead to a wide extension of the disk and the disk may exceed the tidal truncation radius.
To achieve such a condition, an extremely low mass-transfer rate would be a key.
The low mass transferring leads to a very low viscosity in quiescence because of the poor conductivity of the cold disk and resulting decay of magneto-hydrodynamic turbulence \citep{gam98,osa01egcnc}.
If the viscosity is extremely low, mass transferred from the secondary would be stored in a torus at the outer edge of the disk and thus a large amount of matter would be accumulated on the disk at the onset of an outburst.

In the case of $q$ = 0.166(2), the 2:1 resonance radius for a circular orbit, $R_{2:1}/a$ = $(1/2)^{2/3}(1+q)^{-1/3}$, where $a$ is a separation of the binary, is 0.599.
The Roche-lobe radius that is approximated by \citet{egg83rochelobe} is 0.537 with $q$ = 0.166(2).
These values indicate that the 2:1 resonance radius is larger than the Roche-lobe radius.
The distance of $L_1$, the first Lagrangian point, from the primary, which is given by $R_1/a$ = 0.500--0.227$\log{q}$ in \citet{war95book}, is 0.677 and the 2:1 resonance radius seems to be smaller than $L_1$, which may enable the disk to reach the 2:1 resonance radius without colliding with the secondary.
Therefore, the outer edge of the disk perhaps can reach the 2:1 resonance radius depending on conditions.
Another possibility is that the 2:1 resonance can work even if the outer disk does not strictly reach the resonance radius since the 2:1 resonance is very strong.

\subsection{The properties of long-period objects}
\label{longporb}
In \citet{kat15wzsge}, five objects, i.e. V1251 Cyg, RZ Leo, BC UMa, MASTER OT J004527.52+503213.8 (hereafter MASTER J004527) and QY Per, are supposed as candidates of the borderline class between the SU UMa-type DNe and the WZ Sge-type ones or long-period objects\footnote{Although some period bouncers may have long $P_{\rm SH}$, they are not considered here.}.
These objects showed superoutbursts with features similar to those of  WZ Sge-type DNe although these objects have unsuitably long orbital periods for WZ Sge-type ones.
MASTER J004527 and QY Per may be, however, SU UMa-type DNe because no early superhumps were detected in these two objects \citep{Pdot6, Pdot8}.
Except for QY Per, all of these long-period objects showed a single rebrightening as in ASASSN-16eg.

We closely re-examined the past superoutbursts of these long-period objects.
We also reanalyzed the mass ratios of V1251 Cyg \citep{Pdot}, RZ Leo \citep{ish01rzleo} and BC UMa \citep{mae07bcuma} by adding new observations from the AAVSO database and by using a new method proposed in \citet{kat13qfromstageA}.
In these three objects, modulations similar to early superhumps are detected\footnote{In RZ Leo, although \citet{Pdot} suggested that variations of magnitude before the onset of ordinary superhumps are probably early superhumps rather than an extension of ordinary superhumps, \citet{Pdot8} indicated that these modulations may be different from those of typical WZ Sge-type DNe.
However, the existence of ASASSN-16eg supports the existence of long-period WZ Sge-type DNe, and the variations of magnitude similar to early superhumps in RZ Leo may indeed be early superhumps.}.
We summarized our analyses in table \ref{tab:longperiod}.
We list our estimated values for $P_{\rm orb}$ and $P_{\rm stA}$.
The details of the analyses are summarized in the supplementary discussion, figure S1--S10 and table S3--S6.
We should note that these estimations of mass ratios based on the periods of early superhumps and the stage A superhump periods involve large uncertainties mainly because of the short baseline of the data of stage A superhumps.
A reliable value of the period of early superhumps of V1251 Cyg could not be derived from the 2008 superoutburst and thus we could not estimate the mass ratio.
Although the orbital period of BC UMa is longer than those of other WZ Sge-type DNe, it is rather short in comparison with that of RZ Leo or ASASSN-16eg.
The mass ratio of BC UMa also quite small in comparison with that of RZ Leo or ASASSN-16eg.
Thus, this object may be an intermediate class between the SU UMa-type DNe and the WZ Sge-type ones as mentioned in \citet{mae07bcuma}.
The orbital period and mass ratio of RZ Leo are similar to those of ASASSN-16eg.
ASASSN-16eg also may have the same properties as RZ Leo, such as the long recurrence time of superoutbursts.

\subsection{Supercycles of long-period objects}
\label{supercycle}
As we mentioned in subsection \ref{2:1resonance}, the mass-transfer rate may be low in ASASSN-16eg for causing a WZ Sge-type superoutburst.
Similarly, other long-period objects showing WZ Sge-type superoutbursts may have similarly low mass-transfer rates.
The recurrence time of the superoutburst is considered to be proportional to the inverse of the mass-transfer rate \citep{osa95wzsge}.
Therefore, if there is no overlooked superoutburst, the length of supercycle would reflect the mass-transfer rate and the exceptionally long supercycle suggests the exceptionally low mass-transfer rate.

\begin{figure}
\begin{center}
\FigureFile(85mm,95mm){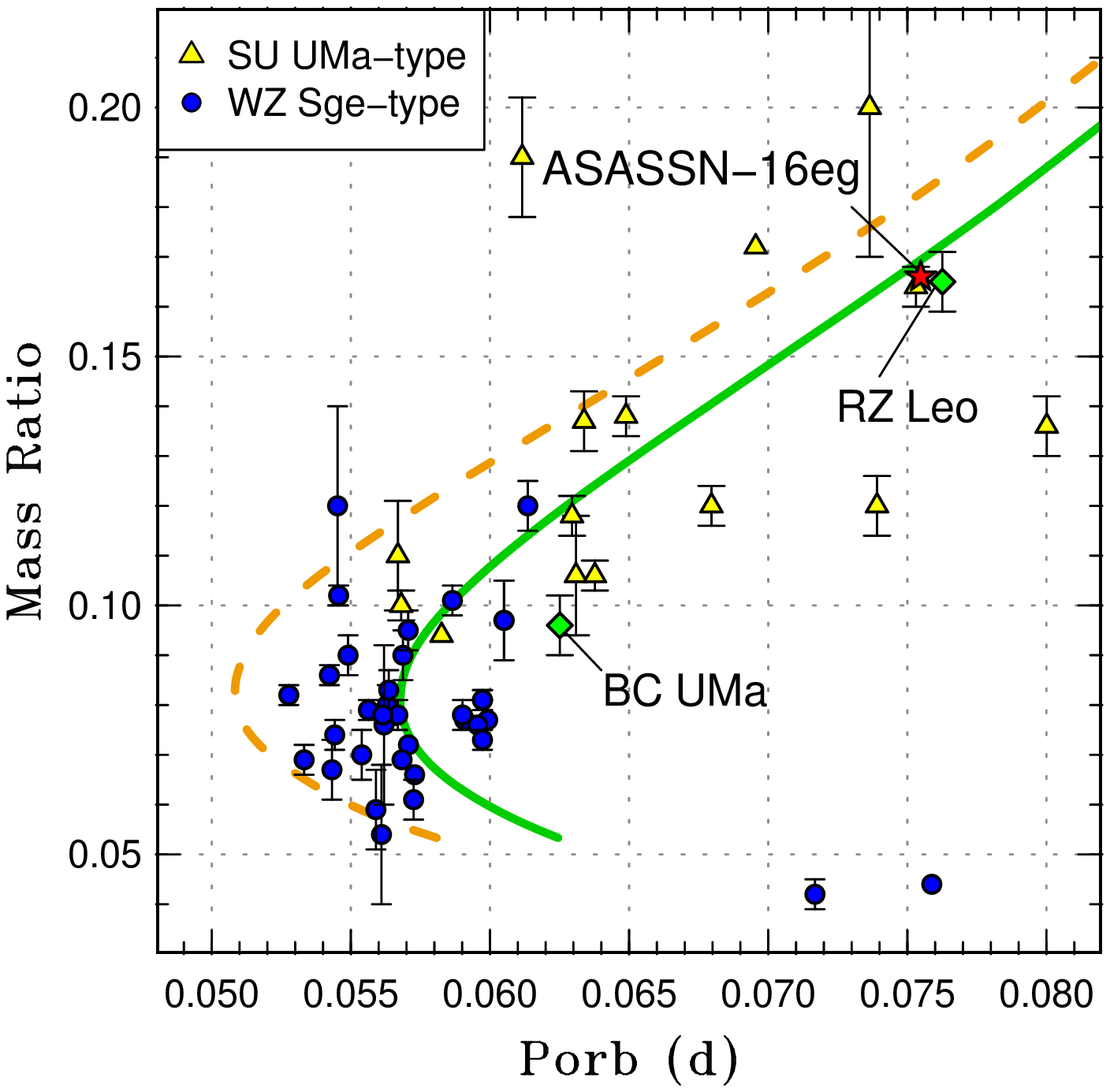}
\end{center}
\caption{Mass ratio, $q$, versus orbital period, $P_{\rm orb}$. The dashed and solid curves represent the standard and optimal evolutionary tracks in \citet{kni11CVdonor}, respectively. The triangle and filled circle represent SU UMa-type and WZ Sge-type DNe, respectively, which are listed in \citet{kat13qfromstageA} and \citet{kat15wzsge}. The star represents ASASSN-16eg. The diamonds represent RZ Leo and BC UMa, respectively. There are two candidates of period bouncers at the lower right in this panel. They have long orbital periods but extremely small mass ratios in comparison with ASASSN-16eg or RZ Leo.}
\label{evol-track}
\end{figure}

We excluded MASTER J004527 from this discussion since this object showed only one superoutburst.
The shortest supercycle of BC UMa is 2 yr as seen in table \ref{tab:longperiod} and this value is rather short for typical WZ Sge-type DNe, which are about decades, although it is long for typical SU UMa-type ones, which are several hundred days.
As \citet{mae07bcuma} mentioned, however, this object may be an intermediate class and seems to be different from other long-period objects (see also figure \ref{evol-track}).
Thus we also excluded this object from this discussion.

The other three objects, V1251 Cyg, RZ Leo and QY Per, have long supercycles comparable to those of other WZ Sge-type DNe.
Although the shortest supercycle of V1251 Cyg is 3 yr and seems to be fairly short, the 1994 superoutburst was not observed well enough \citep{Pdot} and this superoutburst might be one before mass accumulated sufficiently at the onset of the outburst.
These long supercycles indicate that in these long-period objects the mass-transfer rates are low.
From comparison with these long-period objects, it might be indicated that ASASSN-16eg also has a long supercycle and thus a low mass-transfer rate.

We also searched the past outburst of ASASSN-16eg by using Harvard astronomical plate digitalized by Digital Access to a Sky Century @ Harvard (DASCH; \citet{gri09DASCH,Lay10DASCH}) project.
These plates records objects brighter than B $\sim$ 14 -- 17, and could detect the superoutbursts of ASASSN-16eg if this object showed superoutbursts and it was recorded in plates.
There is, however, no record of brightening of ASASSN-16eg and we could not investigate the supercycle of ASASSN-16eg.

\subsection{CV evolution}
\label{CVevolution}
We show the $P_{\rm orb}$ -- $q$ relation in figure \ref{evol-track}, in which we showed SU UMa-type and WZ Sge-type DNe with $q$ values estimated by \citet{kat13qfromstageA} and \citet{kat15wzsge}.
We also included ASASSN-16eg and other long-period objects, RZ Leo and BC UMa.

As described in section \ref{introduction}, the binary separation of CVs becomes shorter because of a loss of the total angular momentum.
Because of Roche overflow, the mass ratio of the binary also reduces.
Thus CVs generally evolve with the mass ratio decreasing and the separation decreasing.
Therefore, it is considered that DNe evolve into WZ Sge-type DNe through SU UMa-type DNe.
This may indicate that the long-period objects including ASASSN-16eg are at an earlier stage of CV evolution than other WZ Sge-type DNe.
However, as mentioned in section \ref{introduction}, the mass-transfer rate decreases as CVs evolve.
In this picture, the mass-transfer rate is considered a function of orbital period.
In cases of long-period objects with long supercycles, the mass-transfer rate appears to be lower than that expected from this picture.
Such systems may be in a state with temporary decreased mass-transfer rate.
However, the mechanism to realize a low mass-transferring state is still unclear.

As one possibility of realizing the temporarily decreased mass-transferring state, we propose the hibernation scenario \citep{Hibernation, liv92hibernation}, which assumes that after a nova eruption, mass transfers from the secondary decreases because of the increase in binary separation and the weakening of irradiation from the primary, and then the binary system would be in the temporally low mass-transferring state.
Although nova-like stars below the period gap are very rare \citep{pat13bklyn}, this hypothesis could explain the reason why the mass-transfer rate can be low even in long-period objects.

Another possibility to explain the low mass-transfer rate in these long-period WZ Sge-type objects is that these objects trace a different evolutionary track from the standard one.
\citet{gol15nuclearevolution} showed that the evolutionary track of CVs depends on the initial conditions of the primary or the secondary.
The objects discussed here, however, appear to be on the standard evolutionary track as judged from the $P_{\rm orb}$ -- $q$ relation (figure \ref{evol-track}), and this possibility appears to be less likely.

\section{Summary}
We report on our photometric observations of the 2016 superoutburst of ASASSN-16eg.
This object showed a WZ Sge-type superoutburst with clear early superhumps and a post-superoutburst rebrightening.
We derived the period of early superhumps to be 0.075478(8) d.
The orbital period, which is almost identical with the period of early superhumps, is exceptionally long for a WZ Sge-type DN.
The mass ratio estimated from the period of stage A superhumps is 0.166(2), which is also very large for a WZ Sge-type DN.
\citet{osa02wzsgehump} proposed that in systems with $q>$ 0.08 the outer edge of the disk cannot reach the 2:1 resonance radius.
However, if the accretion onto the primary proceeds, the disk continues to expand until the resonance radius or tidal truncation radius stops its expansion.
If the mass-transfer rate is low and thus a large amount of matter accumulates on the disk before the onset of an outburst, the outer edge of the disk may reach to or close to the 2:1 resonance radius beyond the tidal truncation radius by violent outburst.

For candidates of long-period objects showing WZ Sge-type superoutbursts, we examined their supercycles, which are considered to reflect the mass-transfer rates.
We found that V1251 Cyg and RZ Leo have long supercycles in comparison to other WZ Sge-type DNe.
This suggests that these long-period objects have low mass-transfer rates.
From comparison to these long-period objects, it is indicated that ASASSN-16eg also has a long supercycle and thus a low mass-transfer rate.

The long orbital period suggests that ASASSN-16eg is in  an earlier stage of CV evolution than other WZ Sge-type DNe.
Although CVs evolve as their mass-transfer rate decreases, long-period objects appear to have a low mass-transfer rate comparable to other WZ Sge-type DNe.
As a mechanism to realize a low mass-transfer rate, we propose the hibernation scenario or possibility that long-period objects trace a different evolutionary track from the standard one.

\section*{Acknowledgement}
This work was supported by a Grant-in-Aid “Initiative for High-Dimensional Data-Driven Science through Deepening of Sparse Modeling” from the Ministry of Education, Culture, Sports, Science and Technology (MEXT) of Japan.
We are grateful to the All-Sky Automated Survey for Supernovae (ASAS-SN) for detecting a large number of DNe and the superoutburst of ASASSN-16eg.
We are thankful to AAVSO and the many amateur observers for providing much of the data used in this research.
This work has made use of data from the European Space Agency (ESA) mission {\it Gaia} (http://www.cosmos.esa.int/gaia), processed by the {\it Gaia} Data Processing and Analysis Consortium (DPAC, http://www.cosmos.esa.int/web/gaia/dpac/consortium).
Funding for the DPAC has been provided by national institutions, in particular the institutions participating in the Gaia Multilateral Agreement.
This work has made use of the VizieR catalogue access tool provided by CDS, Strasbourg, France, and of Astrophysics Data System (ADS) provided by NASA, USA.
This work also has made use of Digital Access to a Sky Century @ Harvard (DASCH) project and we thank Denis Denisenko for his help with use of DASCH.

\section*{Supporting information}
Supplementary discussion, figure S1--S10 and table S1--S6 are reported in the online version.

\bibliography{pasjadd,cvs}

\bibliographystyle{pasjtest1}

\end{document}